\documentclass[usegraphicx,usenatbib]{mn2e}
\usepackage[totalwidth=500pt,totalheight=680pt]{geometry}
\usepackage{times}
\usepackage{textcomp}
\usepackage{amsmath}
\newlength{\colwidth}
\newcommand{\mytilde}{\raise.17ex\hbox{$\scriptstyle\mathtt{\sim}$}}

\title[Coma Cluster far-infrared luminosity function]{A surprising consistency between the far-infrared galaxy luminosity functions of the field and Coma}
\author[S.\ Hickinbottom et al.]
{S. Hickinbottom,$^1$\thanks{E-mail:S.A.Hickinbottom@2011.ljmu.ac.uk}
C. J. Simpson,$^1$
P. A. James,$^1$
E. Ibar,$^2$
D. Carter,$^1$
A. Boselli,$^3$
\newauthor
C. A. Collins,$^1$
J. I. Davies,$^4$
L. Dunne,$^5$
S. Eales,$^4$
C. Fuller,$^4$
B. Mobasher,$^6$
R. F. Peletier,$^7$
\newauthor
S. Phillipps,$^8$
D. J. B. Smith,$^9$
R. J. Smith$^{10}$ and
E. A. Valentijn$^7$\\
$^1$Astrophysics Research Institute, Liverpool John Moores University, IC2, Liverpool Science Park, 146 Brownlow Hill, Liverpool, L3 5RF, UK\\
$^2$Instituto de F\'isica y Astronom\'ia, Universidad de Valpara\'iso, Avda. Gran Breta\~na 1111, Valpara\'iso, Chile\\
$^3$Aix Marseille Universit\'e, CNRS, LAM (Laboratoire d'Astrophysique de Marseille) UMR 7326, 13388, Marseille, France\\
$^4$School of Physics and Astronomy, Cardiff University, The Parade, Cardiff, CF24 3AA, UK\\
$^5$Department of Physics and Astronomy, University of Canterbury, Private Bag 4800, Christchurch 8140, New Zealand\\
$^6$Department of Physics and Astronomy, University of California, Riverside, CA 92521, USA\\
$^7$Kapteyn Astronomical Institute, University of Groningen, Postbus 800, 9700 AV Groningen, The Netherlands\\
$^8$Astrophysics Group, School of Physics, University of Bristol, Bristol BS8 1TL, UK\\
$^9$Centre for Astrophysics, Science \& Technology Research Institute, University of Hertfordshire, Hatfield, Herts, AL10 9AB\\
$^{10}$Department of Physics, Durham University, South Road, Durham DH1 3LE, UK}
\begin{document}

\date{Version of \today}

\pagerange{\pageref{firstpage}--\pageref{lastpage}} \pubyear{2012}

\maketitle

\label{firstpage}

\begin{abstract}

We present new deep images of the Coma Cluster from the ESA \emph{Herschel Space Observatory} at wavelengths of 70, 100 and 160\,$\mu$m, covering an area of $1.75 \times 1.0$ square degrees encompassing the core and southwest infall region. Our data display an excess of sources at flux densities above 100\,mJy compared to blank--field surveys, as expected. We use extensive optical spectroscopy of this region to identify cluster members and hence produce cluster luminosity functions in all three photometric bands. We compare our results to the local field galaxy luminosity function, and the luminosity functions from the \emph{Herschel} Virgo Cluster Survey (HeViCS). We find consistency between the shapes of the Coma and field galaxy luminosity functions at all three wavelengths, however we do not find the same level of agreement with that of the Virgo Cluster.

\end{abstract}


\section{Introduction}
\label{intro}
The study of galaxy clusters is important as they can provide a wealth of information regarding the process of galaxy evolution. It has been well documented that types of galaxies are strongly linked with the nature of their local environments, with a morphology--density relation being found \citep{dre80,whi93,dre97,bal06} such that early-type elliptical and lenticular galaxies are preferentially found in high-density cluster environments.  Several processes have been identified for removing or depleting gas in cluster galaxies, such as ram-pressure stripping \citep{gun72},  `harassment' through multiple high-velocity encounters \citep{moo98}, `starvation' through removal of large-scale weakly-bound gas reservoirs \citep{lar80}, rapid consumption of gas in starbursts resulting from galaxy--galaxy interactions \citep{kee85,jos85,ken87}  or interactions with the core of the cluster mass distribution \citep{mer84,mil86,byr90}. These processes are reviewed by \citet{bos06}. The relative importance of the different processes is likely to be a strong function of local environment, with some only occurring in the cores of the richest clusters, and others being most influential in less extreme environments, and therefore studying a range of environments is of great importance. 

The Coma Cluster is a useful source of information when studying galaxy clusters as it is the richest nearby cluster, at a distance of 100\,Mpc \citep{liu01}. The central core regions are nearly completely virialised \citep{col96} and dominated by early-type galaxies. However, there is clear substructure, the most striking example being a large group centred on NGC\,4839 \citep{neu01}, generally referred to as the `southwest infall region'. Many studies have looked at the star-formation properties of Coma Cluster galaxies, with \citet{ken84} finding several galaxies showing strong H$\alpha$ emission indicating ongoing star formation. This activity was confirmed by \citet{cal93}, \citet{gav91}, \citet{gav98} and \citet{mos05}, although Caldwell et al. found very few such galaxies close to the cluster core. More recently, \citet{smi09,smi12} have studied the importance of environment within the Coma Cluster for the quenching of star formation, showing that it is very important for lower luminosity galaxies but has little effect on the most massive and highly-luminous galaxies.

An additional comparison is provided by the Virgo Cluster, a very well studied galaxy cluster \citep{bos95} which primarily lies at a distance of 17\,Mpc \citep{gav99}. However, through the use of the GOLDMINE database \citet{gav03,gav14} one can see that there is evidence for substructure within the cluster, with groupings at 17, 23 and 32\,Mpc. The Virgo Cluster contains $\sim$2000 optically catalogued galaxies \citep{bin85} of both early and late types. The early types have a velocity dispersion of approximately  589\,km s$^{-1}$, whereas the late types have a velocity dispersion of approximately 700\,km s$^{-1}$ \citep{bin93}. These authors also note that the Virgo Cluster is not fully virialised, and so these values may be affected by substructures or infall velocities. For comparison, the Coma Cluster has a velocity dispersion of ~1008\,km s$^{-1}$ \citep{str99}, and it is a fully virialised environment. The Virgo Cluster has been studied in detail at far-infrared wavelengths as part of the \emph{Herschel} Virgo Cluster Survey (HeViCS; \citealt{dav10}). These observations were carried out using the ESA \emph{Herschel Space Observatory} \citep{pil10}, using the Photodetector Array Camera and Spectrometer (PACS; \citealt{pog10}) at 100 and 160\,$\mu$m, as well as the Spectral and Photometric Imaging Receiver (SPIRE; \citealt{gri10}) at 250, 350 and 500\,$\mu$m.

Far-infrared emission is a sensitive and powerful tracer of the evolutionary state of galaxies. For late-type galaxies, it principally traces star formation activity \citep{lon87,bua96}, with the dominant emission coming from dust thermalisation and re-radiation of energy from high-mass stars. In early-type galaxies, much or all of this emission is instead thought to arise from dust heated by the general radiation field of the older stellar population \citep{lon87,wal96}, and hence far-infrared emission in these galaxies is an indicator primarily of the amount of interstellar medium they have retained. The shape of the far-infrared spectral energy distribution and the total luminosity at these wavelengths thus give insight into both the cold dust content and the star-formation rate \citep{dun00,ken98,ken12}. In turn this information can improve our understanding of galaxy evolution in the cluster environment and the effect of the various galaxy interactions upon the morphology.

Initial results from HeViCS indicated a turnover at faint luminosities \citep{dav10}, which they interpreted as evidence for the stripping process which removes gas and dust being more effective for lower mass objects. One of the motivations for the present study is to determine the far-infrared luminosity function for the Coma Cluster, and to compare it with that for the Virgo Cluster, including the most recent analysis of the HeViCS data \citep{aul13}. Additionally we will investigate how the Coma Cluster luminosity function compares with that found for field galaxies. \citet{bai06,bai09} investigated the Spitzer/MIPS Luminosity Function of the Coma Cluster, and found it to be in agreement with the infrared local field galaxy luminosity function. However, it should be noted that only a small fraction of their galaxies were detected at wavelengths longer than 24\,$\mu$m and therefore the total infrared luminosities were extrapolated using relationships based on optical colours.

Our adopted line of enquiry is to compare cluster galaxies with their field counterparts using the luminosity functions, thus providing an understanding of what effect the local environment has on galaxy luminosity distributions. The likely differences lie in the relative numbers of the different morphological types, as well as a deficiency of gas and dust in the cluster galaxies due to various stripping processes. This study will present results from the deepest far-infrared observations ever of the Coma Cluster, which were obtained using the ESA {\em Herschel Space Observatory}\footnote{Herschel is an ESA space observatory with science instruments provided by European-led Principal Investigator consortia and with important participation from NASA.}. The paper will discuss the multi-wavelength maps of the cluster, and an analysis will be presented of the numbers and luminosities of cluster member galaxies detected at these far-infrared wavelengths. These observations are part of a larger multi-wavelength survey that was instigated with the \emph{Hubble Space Telescope} Advanced Camera for Surveys Coma Cluster Survey (HST/ACS; \citealt{car08}).

The structure of the paper is as follows. In Section 2, the observations and the methodology of the data reduction are described. In Section 3, the analysis of the produced images is discussed, along with the extraction of the sources. In Section 4, the Coma Cluster membership is determined via the use of redshifts. In Section 5, the Coma Cluster luminosity functions are derived and compared to both the field and the Virgo Cluster. A summary of the results is given in Section 6.

\section{Observations and Data Reduction}
\label{obs}

The observations were carried out using the ESA \emph{Herschel Space Observatory}, using the PACS instrument at 70, 100 and 160\,$\mu$m. The observational area was the core of the cluster and the south-west infall region, covering an area of 1.75 by 1.0 degrees. Two scans were performed with simultaneous imaging at 100/160\,$\mu$m and two at 70/160\,$\mu$m, resulting in four separate scans being made. Within each pair, one scan was performed along the long axis of the mapped area, and one along the short axis. The scans were performed at a speed of 20\,arcsec/sec and the total integration time was 27.2 hours, equating to an effective integration time per pixel in the final map of approximately 40 seconds for 70 and 100\,$\mu$m and 80 seconds for 160\,$\mu$m. The OBSID for the scan containing the 70\,$\mu$m maps is 1342224628/9, and for the scan containing the 100\,$\mu$m maps it is 1342233085/6.

The data were reduced using a pipeline written in Jython (a Python implementation written in Java) that was run within the \emph{Herschel} Interactive Pipeline Environment (HIPE; \citealt{ott10}). The pipeline used in this work follows a similar procedure to that described in \citet{iba10}, but using an improved cosmic-ray removal method. The data reduction consists of three main stages: flagging and calibration, deglitching, and imaging. The first stage involves the masking of the bolometers that are known not to be working or are saturated (aided by house-keeping data for the day of the observations). Then the data is astrometrically calibrated by appending the pointing product to each bolometer timeline. The deglitching process involves the removal of all those read-out signals which deviate by more than 5-sigma from the projected map-pixel contributions. The data is projected using {\sc PhotProject} after removing large-scale structure from the timelines using a boxcar high-pass filter (HPF; total width of 1.6\,arcmin). This filtering is essential to remove the dominant $1/f$ noise present in the PACS timelines. The projection uses 3.2\,arcsec pixel sizes for all three wavelengths. The HPF and the map projection is done iteratively in order to mask the read-out contributions to bright pixels before high-pass filtering the timelines, hence removing the sidelobes along the scan directions produced by the boxcar high-pass filter which are clearly seen in the projected images. This process was repeated until the projected map did not differ significantly from the previous iteration. This process results in two different types of sources detected within the map, those which were masked before producing the final version of the map, and those which were not masked and were therefore subject to the high-pass filtering. These two types of sources will be referred to as masked and unmasked sources respectively.

\section{Data Analysis}
\label{data}
The data were analysed using the automatic image detection algorithm \textsc{SExtractor} \citep{ber96}, in order to determine the numbers and positions of sources present in each of the three maps. The values used for the detection threshold, the minimum number of contiguous pixels constituting a detected object, and the deblending contrast parameter were 1.3\,$\sigma$, 6 and 0.01 respectively. An additional flux limit was imposed by requiring that each source as a whole was detected at a five sigma level relative to the rms noise. The background flux level in the final maps is consistent with zero, as is to be expected due to the high-pass filter applied to the timelines before projection. Any sources that were located at the very edges of the maps were discarded due to the increased levels of noise in those regions. 

For each map, there is an optimum aperture size that maximizes the signal-to-noise ratio for point sources. For the 70, 100 and 160\,$\mu$m sources, we determined that the optimum aperture radii are 6, 7 and 10\,arcseconds (Altieri 2013, private communication). The optimum aperture radius requires that the point source fluxes have a corresponding encircled energy fraction correction applied to them. The encircled energy fractions are 63.7, 64.1 and 61.9 per cent for the 70, 100 and 160\,$\mu$m point sources respectively. The final catalogue consists of both masked and unmasked sources, and for unmasked sources this encircled energy fraction will be too large due to the removal of flux during the high pass filter process. Therefore a secondary set of corrections was determined by comparing the final flux of sources with the unprocessed map flux and taking an average of the difference between them. This yielded encircled energy fractions for unmasked point sources of 59.9, 59.7 and 54.4 per cent at 70, 100 and 160\,$\mu$m respectively.

These aperture sizes and encircled energy fractions were also used to calculate the aperture corrected root mean square (rms) noise values, which were found to be 5.7, 6.7 and 7.9mJy at 70, 100 and 160\,$\mu$m respectively.

The fluxes were measured within either the optimum aperture for point sources and corrected by the factors given in Section~\ref{obs}, or a larger aperture for extended sources, where extended is defined as a FWHM greater than $\sqrt{2}$ multiplied by the FWHM of the PACS PSF at the given wavelength. This corresponds to values of 2.47, 2.96 and 4.86 pixels at 70, 100 and 160\,$\mu$m respectively. The larger aperture was defined as having a 12.0 pixel (38.4\,arcseconds) radius for all maps. This value was chosen as it encircles the entirety of the most extended sources that we detect. 

The completeness of the data was determined in order to ascertain how many sources had been missed during the source extraction process. Completeness curves were determined for each wavelength band by inserting 2500 fake sources, 100 at a time, into the maps before running the SExtractor algorithm as described above in order to see how many sources were recovered. This was done for each brightness level between 4\,mJy and 1\,Jy, with intervals of 0.1\,dex.

 The fake source was constructed by stacking a number of bright point sources, and then scaling the flux of the object accordingly. This process was completed twice with two separate fake sources; once for masked sources, and once for unmasked sources. This was done to account for sources that were not masked at any point during the high pass filter process, and as such have a different PSF. The total completeness was determined for each magnitude bin by combining the masked and unmasked completeness using the following equation:

\begin{equation}
C = \frac{c_{m}c_{u}(n_{m}+n_{u})}{c_{u}n_{m}+c_{m}n_{u}}
\label{complete_equation} 
\end{equation}

Here, $c_{m}$ and $c_{u}$ are the masked and unmasked completenesses, and $n_{m}$ and $n_{u}$ are the number of masked and unmasked sources in the bin. This results in a 50 (80)\,per cent completeness of 28.5, 34.5 and 42.0 (34.4, 42.2 and 52.3)\,mJy for the 70, 100 and 160\,$\mu$m maps respectively.

\begin{figure}
\begin{center}
\resizebox{0.95\hsize}{!}{\includegraphics[width=\textwidth]{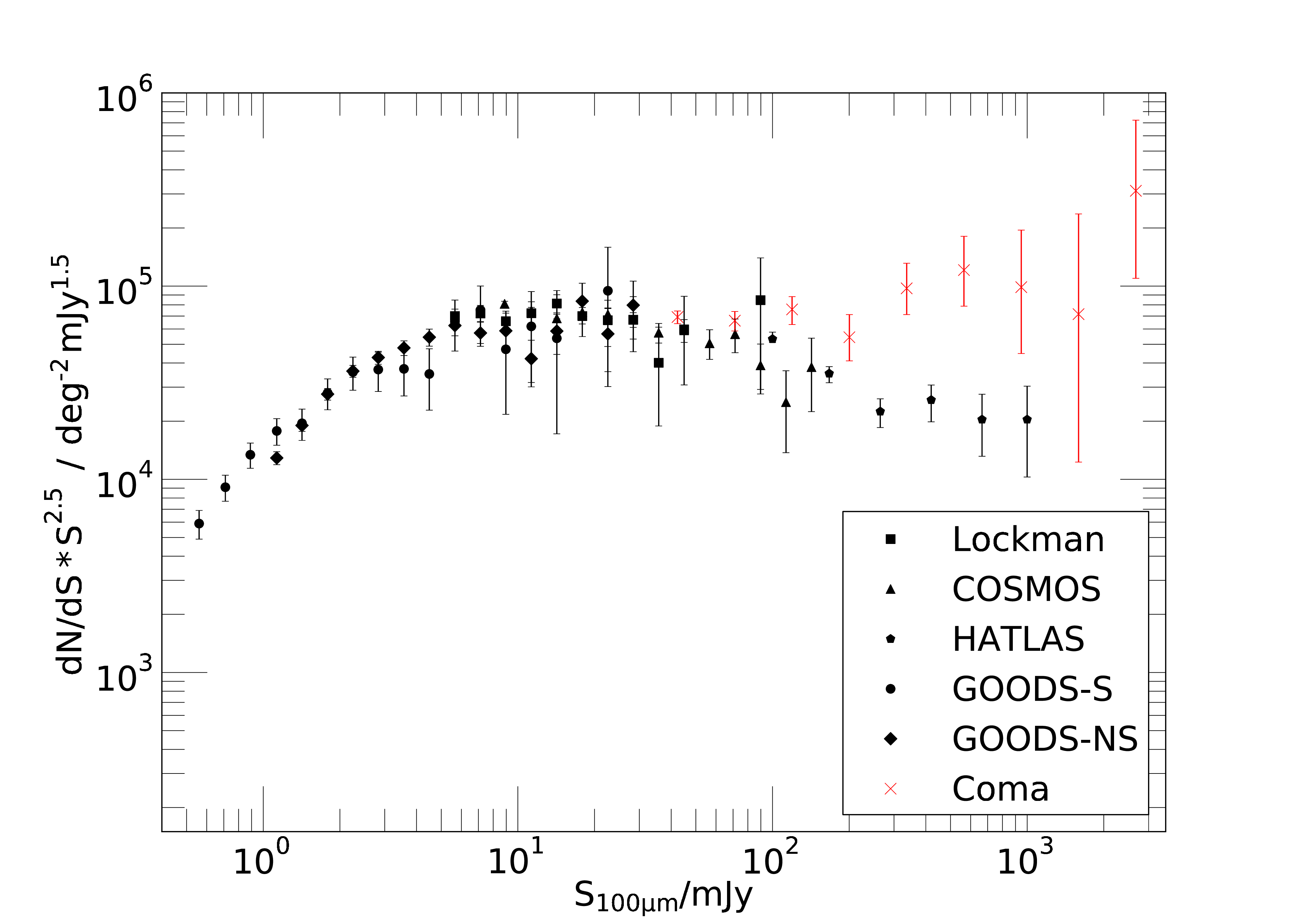}}
\resizebox{0.95\hsize}{!}{\includegraphics[width=\textwidth]{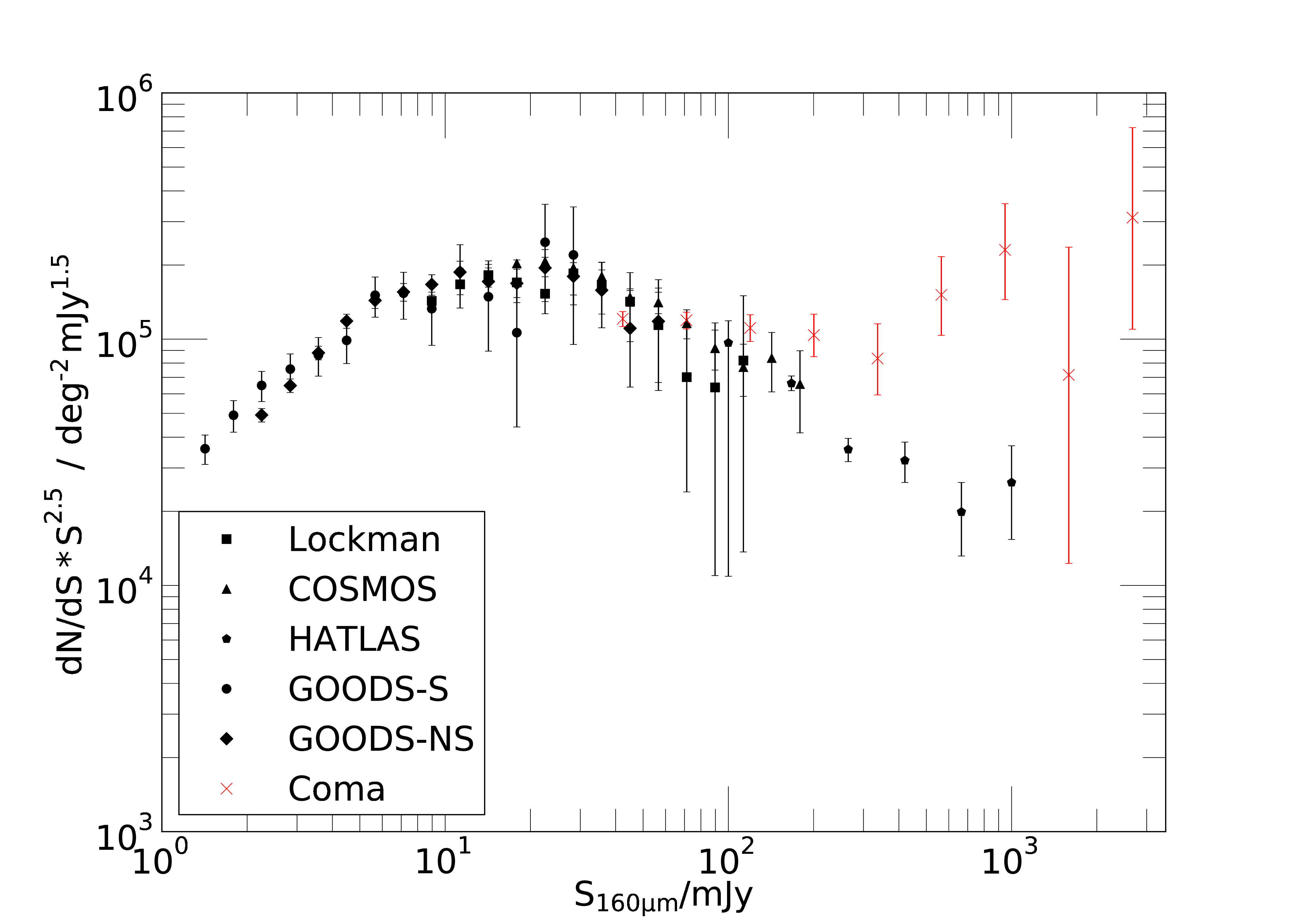}}
\caption{A comparison of \emph{Herschel} Coma source number counts to the H-ATLAS and PEP data at the same wavelength. The top plot shows the 100\,$\mu$m sources, whilst the bottom plot shows the 160\,$\mu$m sources. The data for the Lockman and COSMOS fields were taken from \citet{ber10}. The GOODS field data was taken from \citet{mag13}. GOODS-S refers to the deep scan of the south field only, whereas GOODS-NS is the data from a shallower scan of both the north and south fields.}
\label{pep}
\end{center}
\end{figure}

We detect 201, 370 and 507 sources above the 5\,$\sigma$ limit, in the 70, 100 and 160\,$\mu$m maps respectively. The numbers of these confirmed sources as a function of flux in the 100 and 160\,$\mu$m maps were compared directly to results from the \emph{Herschel} Astrophysical Terahertz Large Area Survey (H-ATLAS: \citealt{eal10}) and the PACS Evolutionary Probe (PEP: \citealt{lut11}), as described in \citet{rig11} and \citet{ber10}. This comparison is shown in Figure~\ref{pep}. The numbers of sources from the Coma data have been corrected for incompleteness, as determined via the method described earlier.  We note the excess of sources at the bright end, which is expected given the presence of a rich galaxy cluster in the mapped area. The faint end of our data is consistent with previous studies to within the respective errors.

The Coma Cluster has previously been observed at far-infrared wavelengths using \textit{IRAS}. The observations found 41 galaxies within 4.2 degrees of the cluster centre, of which 26 were confirmed to be members from their velocities \citep{wan91}. Of these 26 confirmed Coma Cluster members found by \textit{IRAS}, only four lie within our survey region. All four sources match to sources in our \emph{Herschel} catalogue to within 1 arcmin, which is the approximate limiting resolution of the \textit{IRAS} survey. There is a good flux agreement between \emph{Herschel} and \textit{IRAS} for these four sources, however they each have a lower flux as determined by \emph{Herschel}, by 8 to 22 per cent. This is most likely due to the \textit{IRAS} flux being contaminated by other nearby sources due to the large \textit{IRAS} point spread function.

The Coma Cluster was also surveyed to shallower depths at 100 and 160\,$\mu$m by the H-ATLAS survey \citep{eal10}. We matched 153 and 231 sources between the maps to within a distance equal to the optimum aperture radius. We find agreement at the level expected given the measure noise on the maps, down to the detection limit, between the fluxes of these sources down to the 100mJy noise limit of the maps (Smith 2013, private communication).

\section{Cluster Membership}
\label{cluster}
A number of the detected sources in the maps will be foreground or background objects, and have no relation to the Coma Cluster. In order to determine which sources are true Coma Cluster members, the data were compared to the Hectospec Coma Redshift Catalogue.

Data for this catalogue were obtained with the fibre multi-object spectrograph
Hectospec \citep{fab05} at the Multi-Mirror Telescope (now MMT) on the
nights of 2007 April 12--15, with additional queue observations being
made on subsequent nights. The proposal was PA-07A-0260 (PI: Ann
Hornschemeier). These observations used a 270 lines mm$^{-1}$ grating blazed at
$\sim\,5000$\,\AA\ to provide a dispersion of 1.21\,\AA\,pixel$^{-1}$
over a useful wavelength range of 3800--8900\,\AA. A total of 20 fibre
configurations were observed with an integration time of one hour each
and 200 fibres of each configuration were assigned to the redshift
survey.

Targets for the redshift survey were selected from a parent catalogue
of galaxies with Petrosian magnitudes $r<21.3$, within which the
higher priority targets were those with $r<20.3$, those within the
footprint of the \textit{HST} Survey \citep{car08}, within the
\textit{XMM-Newton\/} survey \citep{bri01}, and those
identified with radio sources from the VLA survey \citep{bra00,bra01}. No colour selection criteria were applied.

The data were reduced in a standard manner, and redshifts estimated,
using the Hectospec data reduction pipeline
HSRED\footnote{http://www.astro.princeton.edu/$\mytilde$rcool/hsred/}. A
second redshift estimate was derived for galaxies in the redshift
survey sample using the {\sc iraf} cross-correlation task XCSAO. Each
spectrum was then inspected visually by two separate members of the
HST survey team to resolve discrepancies between HSRED and XCSAO redshifts and
to assess the quality and reliability of the measured redshifts.

The Hectospec catalogue was augmented by redshifts from the NASA/IPAC
extragalactic database\footnote{http://ned.ipac.caltech.edu/} and an
unpublished catalogue by M.M.~Colless and A.M.~Dunn (private
communication). The final catalogue is approximately 90\,per cent complete at $r=19.0$, falling to 50\,per cent at $r=20.3$. Incompleteness is due to the lack of spectra for some galaxies because of fibre proximity constraints, and the
inability to obtain reliable redshifts for the faintest galaxies.

Any source in our \emph{Herschel} catalogue that could be matched to a Hectospec catalogue member, to within a distance equal to the optimum aperture radius, was classified as a true Coma Cluster member; plausible larger values had no effect on the number of matches. From this process it was determined that our data contain 50, 64 and 53 cluster members at 70, 100 and 160\,$\mu$m respectively, with 46 sources being common to all three wavelength bands.

\begin{figure}
\begin{center}
\resizebox{0.95\hsize}{!}{\includegraphics{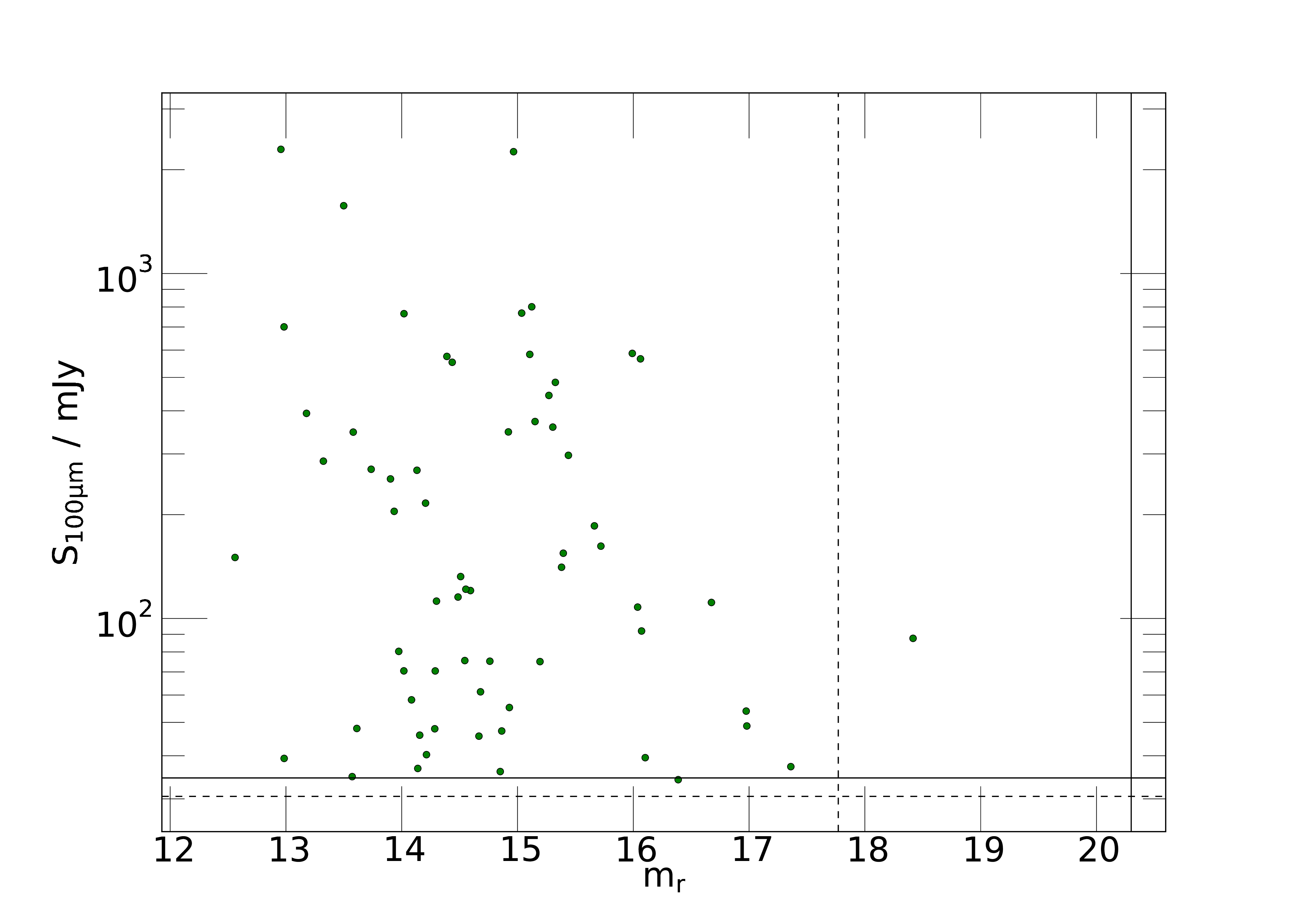}}
\end{center}
\caption{A plot of the 100\,$\mu$m flux against the r band magnitude, for confirmed cluster members. The dashed and solid vertical lines indicate the limiting magnitudes of SDSS (17.77) and Hectospec (20.3) respectively. The dashed and solid horizontal lines indicate the 30 and 50 per cent completeness levels respectively. \citet{bal06b} determine a value of $M_{r}^{*}=-20.49$ for their sample of galaxies, which corresponds to $r=14.51$, therefore as our sources lie around this value, we are not observing the most massive and luminous galaxies.}
\label{rmag}
\end{figure}

Figure~\ref{rmag} shows the infrared flux densities of the 64 sources identified in the 100\,$\mu$m map against their $r$-band magnitudes taken from the Hectospec catalogue. This shows that the limiting magnitude of the Hectospec survey is unlikely to have a significant effect on the number of identifiable sources within our catalogue, and that we can assume we have identified all Coma members within our Herschel sample.

If these confirmed Coma members are removed from the initial sample that was plotted in Figure~\ref{pep}, and then compared to the PEP survey once again, the bright end excess is no longer present, and the trend follows that seen in the PEP counts.

\section{Luminosity Functions}
\label{lumfun}

\subsection{Coma and the Field}
\label{comafield}

\begin{figure}
\begin{center}
\resizebox{0.95\hsize}{!}{\includegraphics{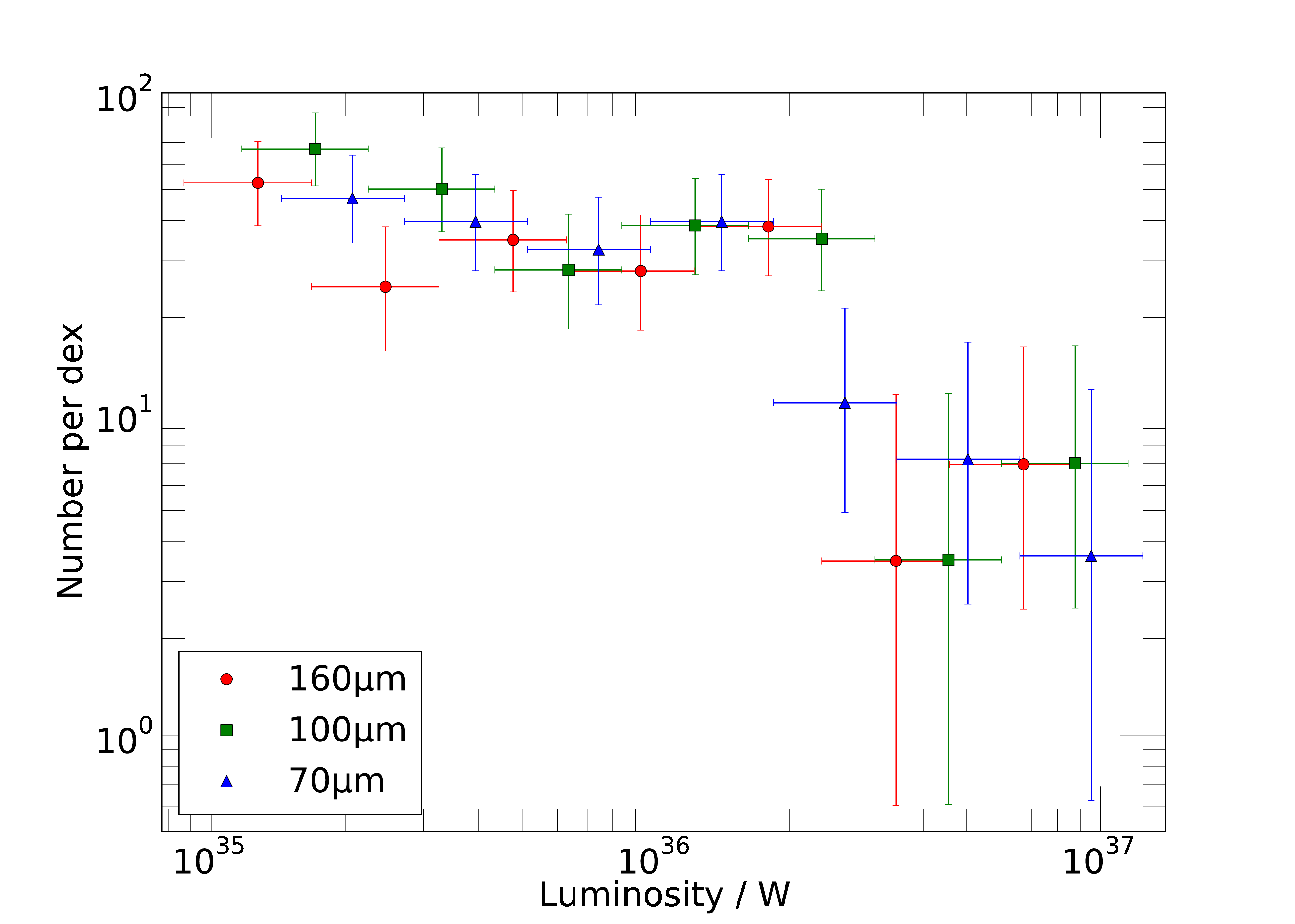}}
\end{center}
\caption[]{Luminosity functions at 70, 100 and 160\,$\mu$m for confirmed Coma Cluster members, shown with blue triangles, green squares and red circles respectively.}
\label{lfs}
\end{figure}

Having produced our catalogues of cluster members, we now construct far-infrared luminosity functions at each wavelength, as shown in Figure~\ref{lfs}. The errors in the numbers were determined via the use of low number statistics, as detailed in \citet{geh86}.

\citet{sch76} showed that galaxy numbers as a function of luminosity can be fitted using three parameters; $\phi^{\star}$, the number of sources, $L^{\star}$, the characteristic luminosity at which a rapid change in the slope of the function is seen, and $\alpha$, a dimensionless parameter which gives the slope of the function at luminosities less than $L^{\star}$. This function is shown in the equation below.

\begin{equation}
\phi(\log{L}) \text{d} \log{L} = \phi^{\star} \ln{10}\left(\frac{L}{L^{\star}}\right)^{\alpha+1}\exp\left(\frac{-L}{L^{\star}}\right)\text{d} \log{L}
\label{scheq} 
\end{equation}

\begin{figure}
\begin{center}
\resizebox{0.95\hsize}{!}{\includegraphics{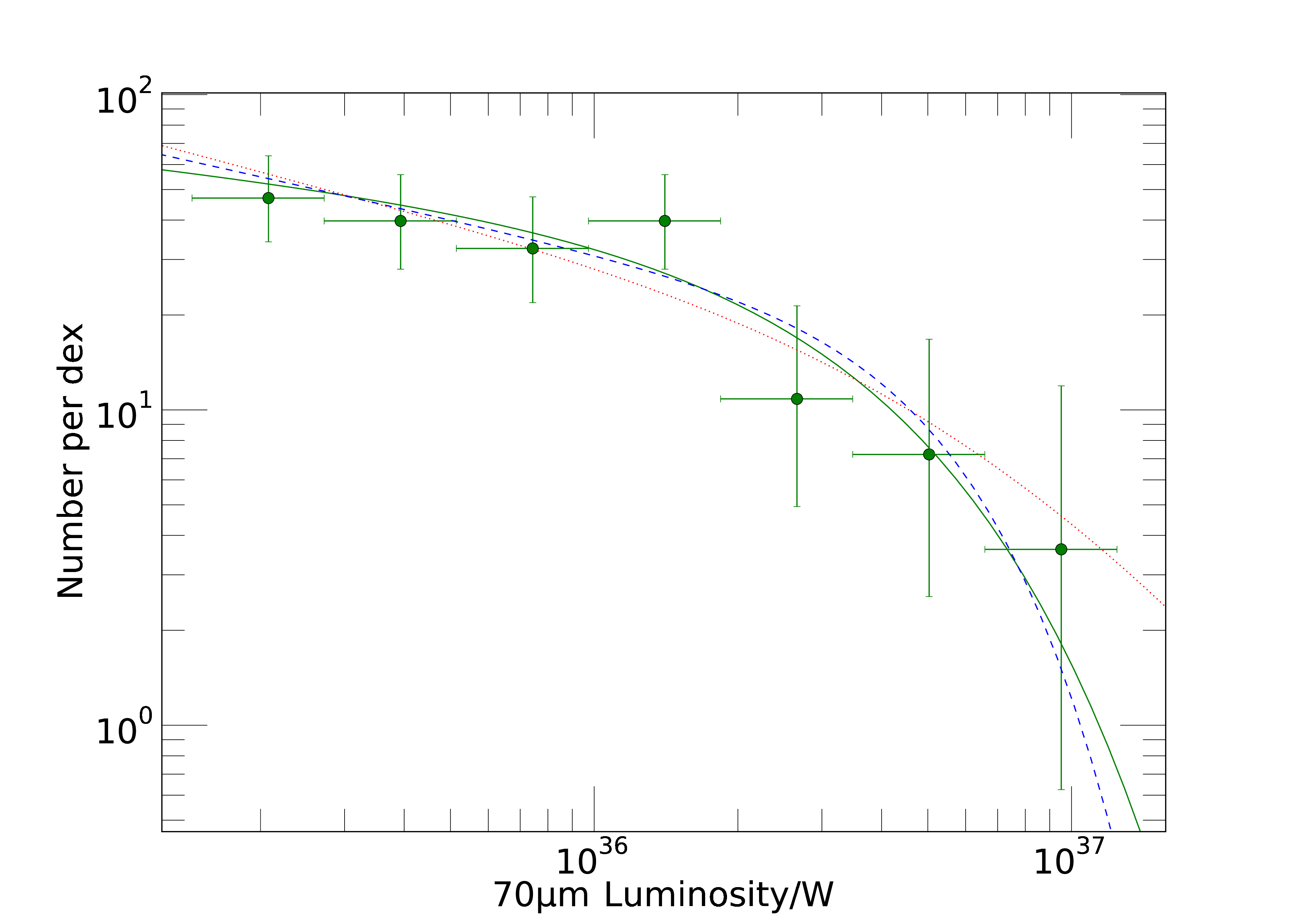}}
\end{center}
\caption[]{A comparison of the Coma and Field 70\,$\mu$m luminosity function. The Coma data are presented as green circles. The green line shows the Schechter function for the Coma data at 70\,$\mu$m, using the parameters presented in Table~\ref{sch}. The blue dashed line shows the luminosity function derived from field galaxies at 60\,$\mu$m as taken from \citet{sau90} and converted to 70\,$\mu$m assuming an M82-like SED. The red dotted line shows the luminosity function as presented by \citet{pat13}, who use the functional form of \citet{sau90}.}
\label{saunderslf}
\end{figure}

We use a maximum likelihood method \citep{mar83} to fit Schechter parameters to the data at each wavelength independently, present the results in Table~\ref{sch} and show these fits in Figures~\ref{saunderslf} and~\ref{hevics}. A Schechter fit was chosen as the extra free parameter in the fits used by \citet{sau90} and \citet{soi87} results in the function being unable to converge on a singular set of values, and produces non-physical results for our data.

\begin{table}
\centering
\begin{tabular}{|c||c|c|c|}
\hline
Wavelength & $\log\phi^{*}$ & $\log L^{*}$ & $\alpha$ \\
 & dex$^{-1}$ & W &  \\ \hline
70\,$\mu$m & $1.19_{-0.37}^{+0.26}$ & $36.53_{-0.22}^{+0.31}$ & $-1.15_{-0.25}^{+0.27}$ \\
\\ 
100\,$\mu$m & $1.07_{-0.36}^{+0.26}$ & $36.70_{-0.22}^{+0.31}$ & $-1.28_{-0.19}^{+0.20}$ \\ 
\\
160\,$\mu$m & $1.28_{-0.28}^{+0.21}$ & $36.42_{-0.19}^{+0.25}$ & $-1.02_{-0.22}^{+0.23}$ \\
\hline
\end{tabular}
\caption{The Schechter parameter values calculated for Coma Cluster galaxies for the three Herschel bands.}
\label{sch}
\end{table}

These data were compared to field galaxy luminosity functions from the literature. In the case of 70 and 160\,$\mu$m there are parametrized luminosity function fits available from the \textit{Spitzer\/} Wide-area Infrared Extragalactic survey (SWIRE; \citealt{pat13}) with an additional dataset at 60\,$\mu$m from \textit{IRAS\/} \citep{sau90}.

The 100\,$\mu$m data were compared to the 90\,$\mu$m binned data presented by \citet{ser04} and \citet{sed11}. Due to the luminosity range covered by these data, and the consequent restriction and rebinning of our own data to match it, the number of Coma members available for the comparison is reduced to a level at which a $\chi^2$ test loses any statistical power. Therefore we fitted a \citet{sau90} function to the data presented in \citet{ser04} using $\chi^2$ minimization, excluding their two lowest bins because they show an unreasonably steep drop in space density that is inconsistent with the local 100\,$\mu$m luminosity function of \citet{row87}, even accounting for the quoted uncertainties. We then compared our data to this parametrized function in the same manner as we have done at the other wavelengths. We choose the \citet{ser04} data because the data in \citet{sed11} do not cover a sufficiently large luminosity range to properly constrain the function we fit.

The Coma data and the various parametrized field luminosity functions are plotted for comparison in Figures~\ref{saunderslf} and \ref{hevics}, with the field galaxy fits having been scaled vertically to fit our data. We run Kolmogorov--Smirnov tests to compare the unbinned luminosity distributions of our cluster members with those predicted by the functions fitted to the field luminosity functions after accounting for completeness. The numbers presented in Table~\ref{KSTable} show the probabilities that our sample is drawn from the same parent distribution as the field luminosity functions. The probabilities for the field comparisons were calculated assuming an effective number of objects using the equation shown below, where $N_{\rm cluster}$ is the number of galaxies in the Coma Cluster and $N_{\rm field}$ is the number of galaxies used to construct the field luminosity function.

\begin{equation}
N_{\rm eff} = \frac{N_{\rm cluster}N_{\rm field}}{(N_{\rm cluster}+N_{\rm field})}
\label{KSeq}
\end{equation}

We find a high probability that Coma is drawn from the same parent distribution as the field luminosity functions. This similarity between the field and Coma luminosity functions is surprising. One would expect the numerous stripping processes as expected from both theoretical and observational studies, would have a large effect in the dense environment of the cluster. A recent study of the Fornax Cluster by \citet{ful14} finds a similar agreement with the field using the PACS 100 and 160\,$\mu$m bands, as well as the SPIRE 250, 350 and 500\,$\mu$m bands. Further analysis will be required in order to provide an explanation for this result. In a future paper we will investigate the masses and morphological types of individual galaxies to look at any underlying reasons for this similarity.

\subsection{Reassessment of Virgo Data}
\label{virgocomparison}

Given the surprising consistency between Coma and the field that suggests the cluster environment has little effect on the far-infrared luminosity function, we reanalyse the data from the Virgo cluster as a means of comparison. \citet{dav10} showed the luminosity functions that had been derived from the HeViCS data. They showed that at all observed wavelengths there was evidence of a turnover at lower luminosities. Further evidence of this was seen with the HeViCS catalogue of bright galaxies \citep{dav12}. \citet{aul13} presented far-infrared fluxes of optically selected Virgo cluster galaxies at 100 and 160\,$\mu$m. We compare these data to the equivalent data in the present study in order to ascertain if Virgo still shows a turnover and for comparison with both our Coma Cluster luminosity function and the relevant field galaxy functions. We use the same Kolmogorov--Smirnov test method to complete these comparisons. These two wavelength bands were used as they are the only wavelengths common to both studies. We follow the assumptions made in \citet{aul13} that all Virgo Cluster galaxies lie at one of three distances, 17, 23 or 32\,Mpc. For the purposes of this study, we use those Virgo sources which lie at 17 and 23\,Mpc and have a measured flux greater than our noise limit. This sample corresponds to that used by the latest HeViCS analysis \citep{dav13}. Analysis of the data shows that the inclusion of those sources at 32\,Mpc, and the assumption that all sources lie at 17\,Mpc, have little effect on the results.

\begin{figure}
\begin{center}
\resizebox{0.95\hsize}{!}{\includegraphics{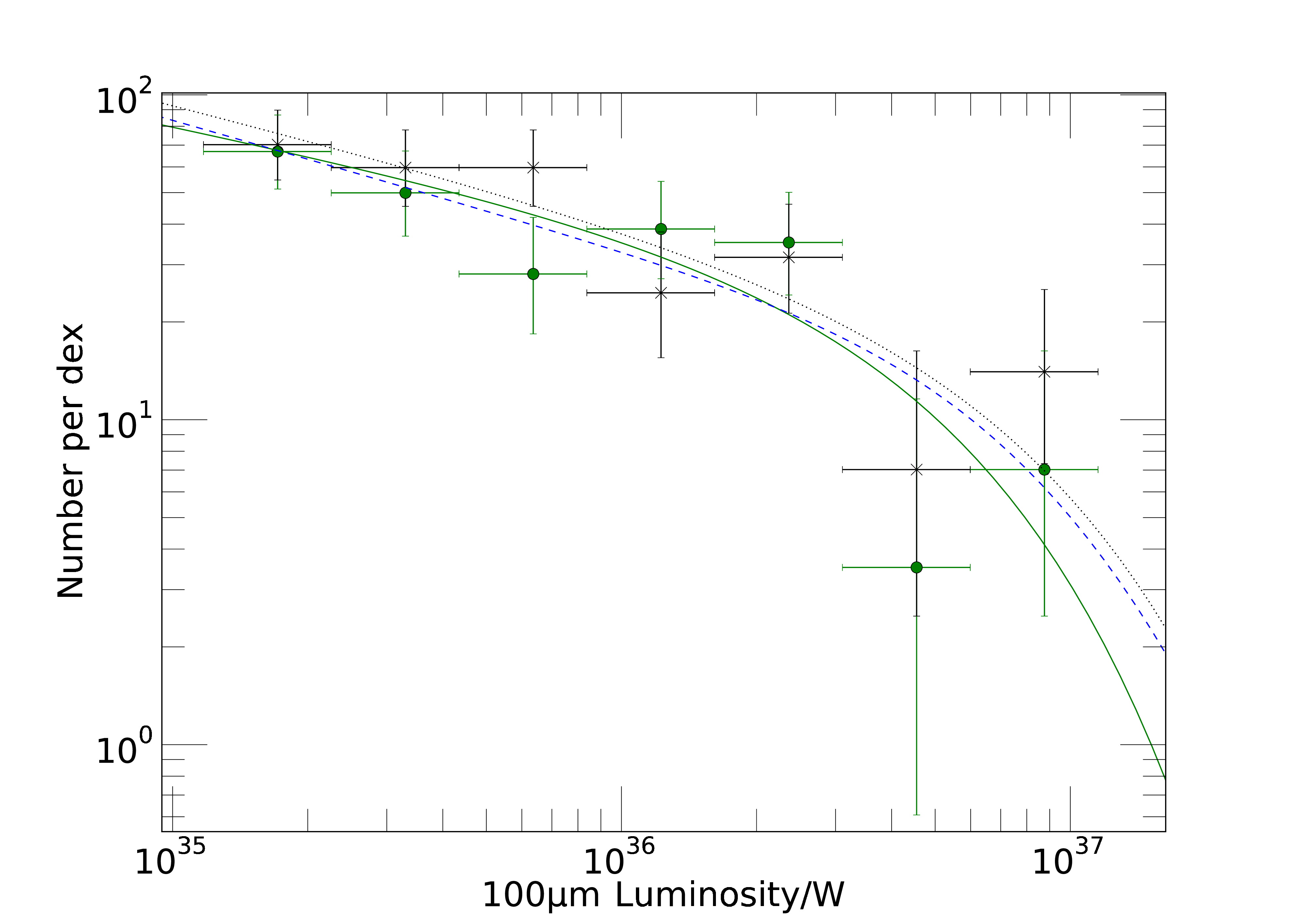}}
\resizebox{0.95\hsize}{!}{\includegraphics{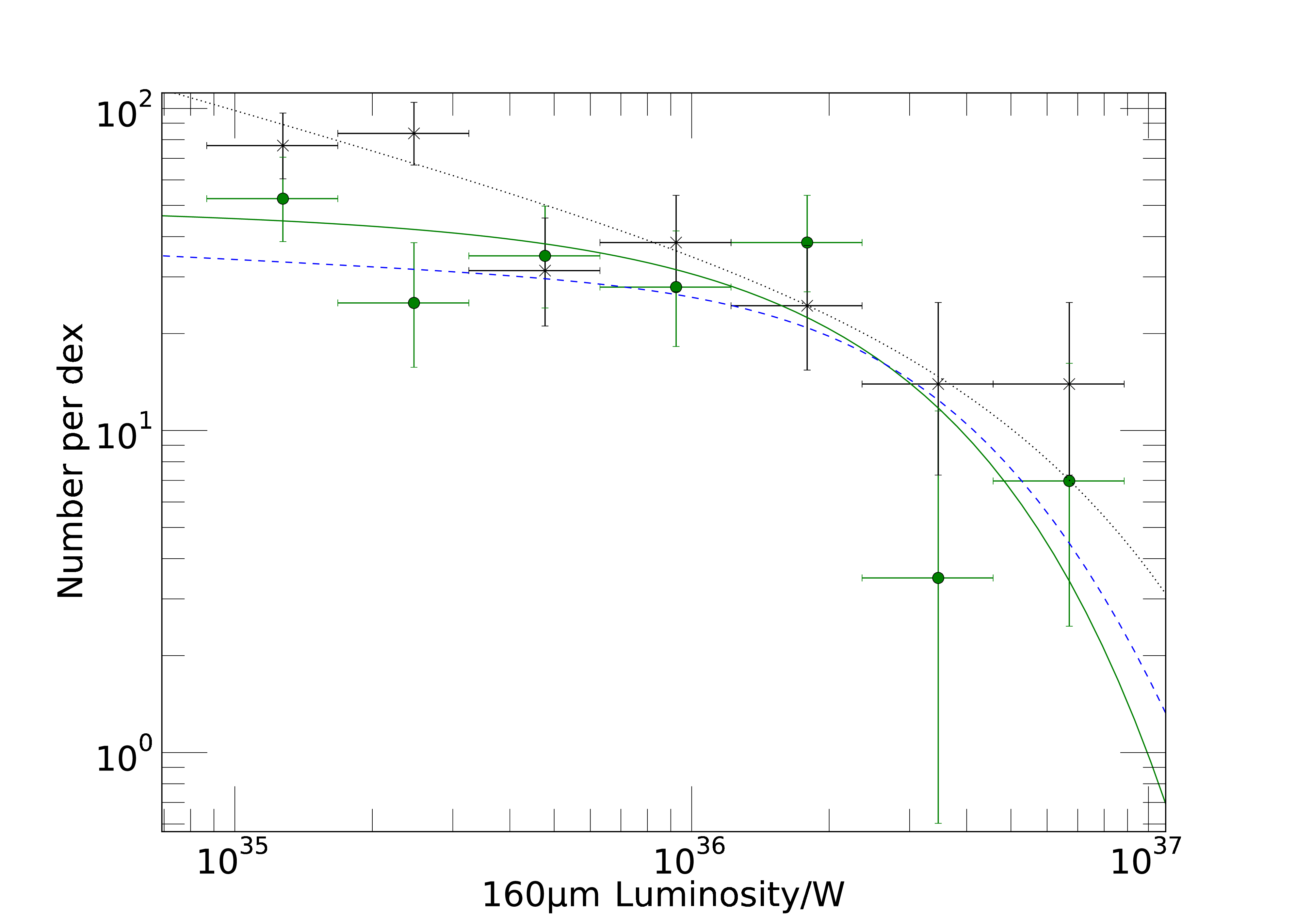}}
\end{center}
\caption[]{A comparison of the Coma, Virgo and Field 100 and 160\,$\mu$m luminosity functions. The Coma data are presented as green circles, and the HeViCS data are presented as black crosses. The green line shows the Schechter function fit for the Coma data, using the parameters presented in Table~\ref{sch}. The black dotted line shows our derived Schechter function fit for the Virgo data of \citet{aul13}. The blue dashed line shows the comparison field luminosity function; \citet{ser04} for the 100\,$\mu$m plot (converted from 90\,$\mu$m assuming an M82-like SED), and \citet{pat13} for the 160\,$\mu$m plot.}
\label{hevics}
\end{figure}

The results of these tests are included in Table~\ref{KSTable}.  The HeViCS data are plotted alongside the present study in Figure~\ref{hevics}, along with a Schechter function fit derived over the range of data included within our comparison. For both Virgo and Coma, we find a clear lack of a turnover at the faint end, in disagreement with the findings of \citet{dav10}. We find a good agreement between the Coma and Virgo Clusters, as well as between the Virgo Cluster and the field, at 100\,$\mu$m. At 160\,$\mu$m however we find no agreement between Virgo and either the Coma Cluster or the field, which still agree with each other reasonably well. We can provide no explanation as to why there would be agreement for one wavelength dataset, but not for the other. However, we note that there is a steeper faint end slope present in the Virgo 160\,$\mu$m data, as can be seen clearly in Figure~\ref{hevics}. This disagreement is especially confusing when one considers that the Virgo Cluster is less virialised than the Coma Cluster, and thus might be expected to have properties intermediate between those of the field and Coma.

\citet{urb11} showed that the Virgo cluster has a $r_{200}$ radius value of 1.08\,Mpc or 3.9\textdegree\ at their adopted distance of 16.1\,Mpc, corresponding to a value of 1.16\,Mpc at our defined distance of 17.0\,Mpc. \citet{arn05} showed that the $r_{200}$ radius value can be adopted as the virial radius for Virgo, even though the cluster is not virialised. The HeViCS observations therefore cover an area out to over twice the virial radius. In comparison the virial radius of the Coma Cluster is 2.9\,Mpc or 1.7\textdegree\ \citep{lok03}, with our observations covering an area out to approximately this distance. The Virgo observations therefore cover a greater area outside of the core environment relative to the Coma observations. Future work will take this factor into account by looking at subsets of the Virgo observations and comparing similar environments of the clusters.

\begin{table}
\centering
\begin{tabular}{|c||c|c|c|c|}
\hline
Wavelength & Field LF & Coma/Field & Coma/Virgo & Virgo/Field \\\hline
70\,$\mu$m & Patel & 73 & &  \\
\\ 
70\,$\mu$m & Saunders & 78 & &  \\
\\ 
100\,$\mu$m & Serjeant & 86 & 94 & 99 \\ 
\\
160\,$\mu$m & Patel & 62 & 9 & 0.6  \\
\hline
\end{tabular}
\caption{The percentage results of Kolmogorov--Smirnov tests to compare the luminosity distribution of the luminosity functions of Coma, Virgo and the field. The \citet{sau90} 60\,$\mu$m luminosity function has been converted to 70\,$\mu$m, and the \citet{ser04} 90\,$\mu$m luminosity function to 100\,$\mu$m assuming the SED of M82 \citep{sil98}.}
\label{KSTable}
\end{table}

\section{Summary}
We have presented the deepest far-infrared observations to date of the Coma Cluster of galaxies. The observations are used to produce far-infrared number counts within this area, and in combination with an optical redshift catalogue are used to derive far-infrared luminosity functions at the three \emph{Herschel}/PACS wavelengths (70, 100 and 160\,$\mu$m) for the Coma Cluster. These functions are compared with both the field galaxy luminosity function and similar results from the HeViCS survey of the Virgo Cluster. We see that our Coma Cluster luminosity functions have shapes surprisingly consistent with those of the field galaxies at the three far-infrared wavelengths we probe. The Coma Cluster luminosity function also matches well with the newly derived luminosity function for the Virgo Cluster at 100\,$\mu$m, which no longer shows evidence of a faint-end turnover. The same consistency is not seen with the Virgo Cluster at 160\,$\mu$m, but this cannot be easily explained. We note that it is not due to our sample consisting only of the most luminous galaxies, which would be unaffected by stripping, as Figure~\ref{rmag} shows that our sources lie around the value of $L^{*}$ for the $r$-band. Future work will look in detail at the dust properties and stellar masses of the detected Coma Cluster members to try to identify a mechanism or process that could result in such similarity between the cluster and field environments.

\section*{Acknowledgements}

Scot Hickinbottom acknowledges a STFC studentship.

We acknowledge Mustapha Mouhcine for his help in the initial preparation of this study, Matthew W. L. Smith for providing the H-ATLAS data for the source flux comparison, and the anonymous referee for some of his helpful comments.

\emph{Herschel} is an ESA space observatory with science instruments provided by European-led Principal Investigator consortia and with important participation from NASA. PACS has been developed by a consortium of institutes led by MPE (Germany) and including UVIE (Austria); KU Leuven, CSL, IMEC (Belgium); CEA, LAM (France); MPIA (Germany); INAF-IFSI/OAA/OAP/OAT, LENS, SISSA (Italy); IAC (Spain). This development has been supported by the funding agencies BMVIT (Austria), ESA-PRODEX (Belgium), CEA/CNES (France), DLR (Germany), ASI/INAF (Italy), and CICYT/MCYT (Spain). HIPE is a joint development by the Herschel Science Ground Segment Consortium, consisting of ESA, the NASA Herschel Science Center and the HIFI, PACS and SPIRE consortia.

This research has made use of the NASA/IPAC Extragalactic Database (NED) which is operated by the Jet Propulsion Laboratory, California Institute of Technology, under contract with the National Aeronautics and Space Administration.

\label{lastpage}

\end{document}